\documentclass{article}
\usepackage{vmargin}

\setpapersize{A4}
\setmargins{2.5cm}       % margen izquierdo
{1.5cm}                  % margen superior
{16.5cm}                 % anchura del texto
{23.42cm}                % altura del texto
{10pt}                   % altura de los encabezados
{1cm}                    % espacio entre el texto y los encabezados
{0pt}                    % altura del pie de página
{2cm}                    % espacio entre el texto y el pie de página
\usepackage[utf8x]{inputenc}
\usepackage{ucs}
\usepackage{amsmath}
\usepackage{amsfonts}
\usepackage{amssymb}
\usepackage{appendix}
\usepackage{hyperref}
\usepackage{makeidx}
\usepackage{graphicx}
\usepackage{amsthm}
\usepackage{color}
\usepackage{comment}
\usepackage{rotating}
\usepackage{pictex}
\usepackage{makeidx}
\usepackage{float}
\usepackage{wrapfig}
\usepackage{epsfig}
\usepackage[affil-it]{authblk}

\title{The confined helium atom; an Informational approach}

\author[1]{C. R. Esta\~{n}\'on\footnote{carloscbiuam1@gmail.com}}
\author[2]{H. E. Montgomery Jr\footnote{ed.montgomery.jr@gmail.com}}
\author[3]{J. C. Angulo\footnote{angulo@ugr.es}}
\author[4]{N. Aquino\footnote{naa@xanum.uam.mx}}

\affil[1,4]{Departamento de  F\'isica, Universidad Aut\'onoma Metropolitana-Iztapalapa,
Av. San Rafael Atlixco $186$, Col. Vicentina, CP $09340$ CDMX, M\'exico.}
\affil[2]{Chemistry Program, Centre College, 600 West Walnut Street, Danville, KY 40422-1394, USA.}
\affil[3]{Departamento de F\'{\i}sica At\'{o}mica, Molecular y Nuclear, Universidad de Granada, Granada 18071, Spain, and Instituto Carlos I de F\'{\i}sica Te\'orica y Computacional, Universidad de Granada, Granada 18071, Spain.}

\begin{document}

\maketitle

\begin{abstract}
    In this work we study the helium atom confined in a spherical impenetrable cavity by using informational entropies. We use the variational method to obtain the energies and wave functions of the confined helium atom as a function of the cavity radius $r_0$. As trial wave functions we use one uncorrelated function and four functions with different degrees of electronic correlation. We computed the Shannon entropy, Fisher information, Kullback--Leibler entropy, Disequilibrium, Tsallis entropy and Fisher--Shannon complexity, as a function of the box radius $r_0$. We found that these entropic measures are sensitive to electronic correlation and can be used to measure it. These entropic measures are less sensitive to electron correlation in the strong confinement regime ($r_0<1$ a.u.).
%\keywords{Confined helium atom, Tsallis entropy, Fisher-Shannon complexyty measure}
\end{abstract}

%========================
\section{Introduction}
%========================
Spatially confined quantum systems have become the subject of increasing attention because of the wide variety of problems in physics and chemistry that can be modelled through them. Some of these problems are atoms trapped in cavities, in zeolite channels, in fullerenes, the electronic structure of atoms and molecules subjected to high external pressures, the behavior of the specific heat of a monocrystal solid under high pressure \cite{Froman87,Jasko96,Buch,Conne,Sabin2009,Sen2009,Sen2014,LeyKoo2018}, etc. This growing interest is also due to the fabrication of quantum systems of nanometric sizes with potential technological applications such as in quantum wires, dots and wells \cite{Banyai,Harrison}.\\

In 1937 Michels et. al. \cite{Michels} studied the variation of the polarizability of the hydrogen atom subjected to high external pressures. They proposed a model in which a hydrogen atom is confined inside an impenetrable spherical cavity with the nucleus clamped in the center of a sphere of radius $r_0$. This model is known as the confined hydrogen atom (CHA) \cite{Ley1,Marin,Aquino2009,Sen2,Ed1}. Furthemore has been very successful, and electronic properties of multi-electron atoms have been studied with it. This model has been extended to use cavities of shapes other than spherical to study also atoms and molecules trapped inside cavities. \\

Confined helium--like atoms are the simplest confined many-electron atoms, consisting of a nucleus with nuclear charge $Z$ and two electrons. For this type of systems of two electrons, different methods have been developed  from time-independent perturbation theory up to Quantum Monte Carlo methods \cite{Aquino2009,tenseldam,Gimarc,Lude1,Luden2,Joslin,Aquino2003,Aquino2006,Antonio2008,Antonio2010,Laughlin,Ed2010,Wilson2010,Aquino2014,Lesech,Bhatta,Jgo2016,Doma}. Most of these studies have been devoted to the calculation of the ground state energy as a function of the confining radius. Recent work has addressed the correlation energy due to the radial and angular contribution of trial wave functions \cite{Wilson2010,Aquino2014}. Subsequent work has been devoted to the study of low energy excited states.\\

Information theory has been used in the study of one-, two- and three-dimensional (1D, 2D and 3D) systems \cite{Ben2006,Martins2008,Martins2009,Salazar2021,Laguna2022,Salazar2022,Salazar2023}, in free systems and in systems subject to spatial confinement. As examples, we can mention the study of free and spatially confined hydrogen and helium atoms \cite{Cesar,Yanes1994,Sen2007,Aquino2013}.

As discussed in \cite{KDSen2011}, in recent years a variety of complexity measures have been defined and applied to the study of physical, biological, mathematical, computer science, etc. systems. These quantities are obtained from probability density and expectation values,\cite{HoSmith,Ho1998,Gadre1984,Gadre1985,Gadre1985b,Sen2007}.
Complexity measures are quantities that are well-defined in any conjugate space.\newline

Each complexity measure is directly related to information entropies and this allows us to determine global or local features of the probability density; for example: a global-local measure is described by the Fisher-Shannon complexity which relates the Shannon entropic power to the Fisher entropy, with the Fisher entropy providing a local measure of the probability density and the Shannon entropic power providing a global measure.\newline

The work is organized as follows:  In section 2 we briefly describe the confined helium system, its solution by the variational method and the informational measures used in this work: the Shannon entropy, Fisher information, Tsallis and Kullback--Leibler entropies, the disequilibrium and the Fisher-Shannon complexity. In section 3 we discuss our results.  Finally, in section 4 we give our conclusions.

%==================================================================
\section{Theoretical background}
%=========================================================================

%=========================================================================
\subsection{Confined helium atom; ground state energy, wavefunctions and its probability densities}
%==================================================================
The Hamiltonian of a helium--like atom confined in an impenetrable spherical box of radius $r_0$ (in the infinite nuclear mass approximation), in atomic units ($\hbar=e=m_e=1$), is given by:

\begin{equation}\label{eq:hamiltoniano}
    \hat{H}=-\frac{1}{2}\nabla_1^2-\frac{1}{2}\nabla_2^2+V(\vec{r}_1,\vec{r}_2),
\end{equation}
where the first two terms on the right-hand side are the electron kinetic energies, and the potential energy is given by:

\begin{equation}
 V(\vec{r}_1,\vec{r}_2)=\begin{cases}-\frac{Z}{r_1}-\frac{Z}{r_2}+\frac{1}{r_{12}}, & r_1,r_2< r_0\\ 
 \infty, \quad \text{while} & r_1\geq r_0 \quad \text{or} \quad r_2 \geq r_0\end{cases},
\end{equation}

where $r_1$ is the distance from the nucleus to electron 1, $r_2$ is the distance from the nucleus to electron 2, $r_{12}=|\vec{r}_1-\vec{r}_2|$ is the distance from electron 1 to electron 2 and $Z=2$ is the nuclear charge for the helium atom. Inside the spherical barrier $r_1,r_2< r_0$, the potential energy is formed by the Coulombic attractive interaction between the electrons and the nucleus, and the repulsive interaction between the electrons. \\

In order to solve the problem of finding the energy eigenvalues it is convinient to define the Hylleraas coordinates: $s \equiv r_1+r_2$, $t \equiv -r_1+r_2$ and $u \equiv r_{12}$. The Hamiltonian of the confined helium atom, in the sphere, in Hylleraas coordinates can be written as:

\begin{equation}
    \begin{split}
        \hat{H}= & -   \left(\frac{\partial^2}{\partial s^2} + \frac{\partial^2}{\partial t^2}  +\frac{\partial^2}{\partial u^2} \right)- 2 \frac{s(u^2-t^2)}{u(s^2-t^2)} \frac{\partial^2}{\partial s \partial u} \\
         & - 2 \frac{t(s^2-u^2)}{u(s^2-t^2)} \frac{\partial^2}{\partial t \partial u}-\frac{4s}{(s^2-t^2)} \frac{\partial}{\partial s}+ \frac{4t}{(s^2-t^2)} \frac{\partial}{\partial t} \\
         & -  \frac{2}{u} \frac{\partial}{u \partial u}-4Z \frac{s}{s^2-t^2}+ \frac{1}{u}.
    \end{split}
\label{eq:HHy}
\end{equation}

In this report we only study the ground state of the confined helium atom. To obtain the approximate energy and its corresponding wave function, as a function of the box size $r_0$, we use the variational method. We propose two types of trial wave functions: uncorrelated and correlated wave functions.
 
%=====================================================================================================
\subsection*{Uncorrelated wave function}
%=====================================================================================================
According to the direct variational method the wave function is constructed as the wave function of the free (unconfined) system times a cut-off function. The simplest wave function is given by the product of two hydrogen--like wave functions, multiplied by the cut-off function $(r_0-r_1)(r_0-r_2)$ that makes the wave function vanish at the confining surface of the spherical cavity. The uncorrelated wave function is the following:

\begin{equation}
    \psi_0=B e^{-\alpha (r_1 + r_2)}(r_0-r_1)(r_0-r_2),
\end{equation}
which in Hylleraas coordinates it can be written as: 

\begin{equation}
    \psi_0(s,t,u)=B e^{-\alpha s}\left(r_0-\frac{s-t}{2}\right)\left(r_0-\frac{s+t}{2}\right),
\end{equation}

where $\alpha$ is a variational parameter.

%=====================================================================================================
\subsection*{Wave functions with electronic correlation}
%=====================================================================================================
We used four wave functions that include electronic correlation. The trial wave functions in Hylleraas coordinates are the following:

\begin{equation}\label{eqpsi1}
    \psi_1(s,t,u)=B e^{-\alpha s}(1+\beta u)\chi(s,t,u;r_0),
\end{equation}

\begin{equation}\label{eqpsi2}
    \psi_2(s,t,u)=B e^{-\alpha s}(1+\beta u+\gamma t^2)\chi(s,t,u;r_0),
\end{equation}

\begin{equation}\label{eqpsi3}
    \psi_3(s,t,u)=B e^{-\alpha s}(1+\beta u+\gamma t^2+\delta s^2)\chi(s,t,u;r_0),
\end{equation}

\begin{equation}\label{eqpsi4}
    \psi_4(s,t,u)=B \sum_{n, m, \ell}^2 C_{nlm} e^{-\alpha s}s^n t^m u^\ell \chi(s,t,u;r_0),
\end{equation}

where $n+m+l \leq 2$, $\chi(s,t,u;r_0)=\left(r_0-\frac{s-t}{2}\right)\left(r_0-\frac{s+t}{2}\right)$ is the cut-off function, and $\alpha, \beta, \gamma$, $\delta$ and $C_{nlm}$ are variational parameters.

\begin{table}[h!]
\caption{The ground state energy for uncorrelated and correlated wave functions as a function of confinement radii $r_0$.}
\centering
 \begin{tabular}{ c c c c c c} 
\hline
$r_0$(a.u.) & $E(\psi_0$) & $E(\psi_1)$ & $E(\psi_2)$  & $E(\psi_3)$ & $E(\psi_{4})$ \\ [0.5ex]
 \hline\hline
0.5000  & 22.9229  & 22.9043 & 22.8321 &  22.7765 & 22.7426  \\  
0.6000  & 13.4250  & 13.3986 & 13.3645 &  13.3421 & 13.3204  \\  
0.7000  & 7.9968   & 7.9642  & 7.9490 &  7.9382  & 7.9278  \\  
0.8000  & 4.6656   & 4.6282  & 4.6224 &  4.6201  & 4.6120  \\  
0.9000  & 2.5117   & 2.4706  & 2.4691 &  2.4706  & 2.4642  \\  
1.0000  & 1.0625   & 1.0186  & 1.0185 &  1.0214  & 1.0172  \\
2.0000  & -2.5284 & -2.5797 & -2.5976 &  -2.5994 & -2.5977  \\
3.0000  & -2.7935 & -2.8419 & -2.8651 &  -2.8659 & -2.8679  \\
4.0000  & -2.8301 & -2.8763 & -2.8955 &  -2.8960 & -2.8981  \\  
5.0000  & -2.8391 & -2.8843 & -2.9003 &  -2.9007 & -2.9023  \\  
6.0000  & -2.8425 & -2.8871 & -2.9015 &  -2.9018 & -2.9029  \\  
10.0000 & -2.8462 & -2.8900 & -2.9022 &  -2.9026 & -2.9033  \\  
$\infty$ & -2.84766  & -2.8911 & -2.9024 & -2.9027 & -2.9036 \\[1ex]  
 \hline
\end{tabular}
\label{tabla:alfacero}
\end{table}

%====================================================================
\subsubsection*{Energy calculations}
%====================================================================
As we mentioned above we use the variational method to obtain the approximate energy and wave functions, in this approach we minimize the energy functional 

\begin{equation}
  E=\frac{\langle \psi | \hat{H} |\psi \rangle}{\langle \psi | \psi \rangle},
\end{equation}

with respect to the variational parameters, where $\hat{H}$ is the Hamiltonian in Hylleraas coordinates (eq. \eqref{eq:HHy}) and $\psi=\psi_i, i=0,...,4$. 

\begin{figure}[h!]
    \centering
    \includegraphics[width=0.8\textwidth]{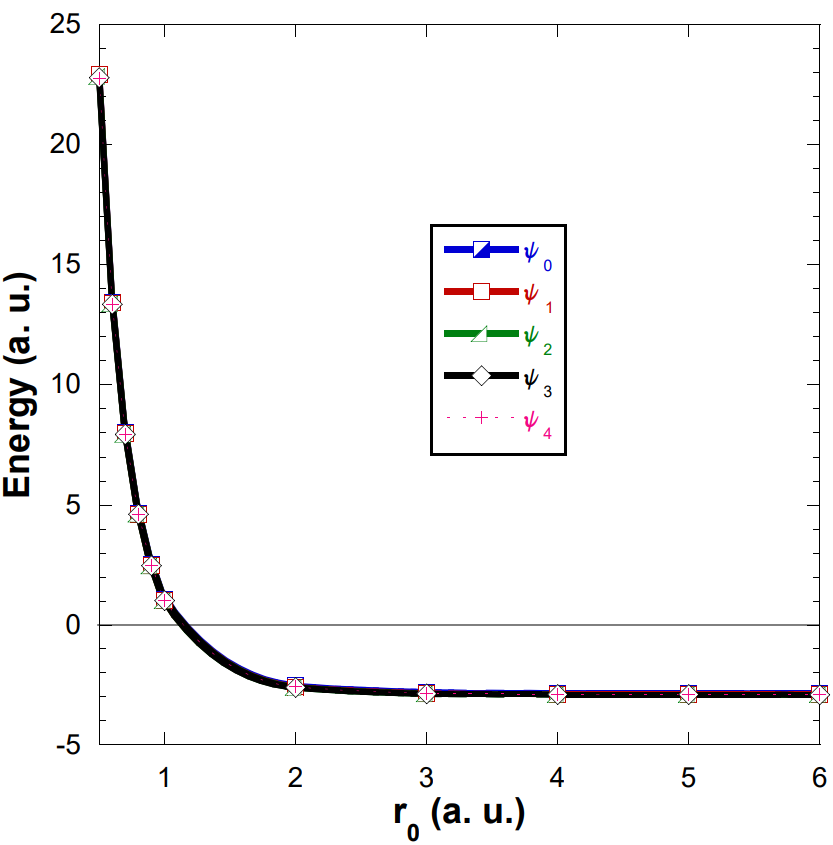}
    \caption{Energy variation for the helium atom with and without electronic correlation, varying the confinement radius $r_0$.}
    \label{Fig:EnergiaCC}
\end{figure}

For the confined helium atom, different expressions \cite{tenseldam, Pan2003, Nascimento2020} have been used to evaluate the integrals involved in the energy functional; those expressions are equivalent and provide the same results. The expression that we used to evaluate the integrals in the energy functional is the following \cite{Pan2003}:

\begin{equation}
    \begin{split}
        \int f \text{d}\tau&=2\pi^2\int_0^{r_0}\text{d}s\int_0^{s}\text{d}t\int_t^{s}f(s,t,u)(s^2-t^2)u\text{d}u\\
        &+2\pi^2\int_{r_0}^{2r_0}\text{d}s\int_0^{2r_0-s}\text{d}t\int_t^{s}f(s,t,u)(s^2-t^2)u\text{d}u
    \end{split}
\end{equation}

where $f$ can be either $\hat{H}$ or the probability density $|\psi|^2$. In this way we proceed to numerically evaluate these expressions, the results can be seen in Table \ref{tabla:alfacero}, as well as graphically in the Figure \ref{Fig:EnergiaCC}.

%=========================================================================
\subsubsection*{Quantum probability density}
%=========================================================================

The one electron probability density is obtained by integrating over the coordinates of the other electron. The probability density associated with the wave function $\psi_0$ is given by

\begin{equation}
    \rho_0(\Vec{r})= B e^{-2 \alpha r}
\end{equation}

whereas for the wave functions $\psi_1, \cdots \psi_4$ the probability density is obtained by:

\begin{equation}\label{eq:densidadC}
    \begin{split}
        \rho_{i}(\vec{r})=&\frac{2\pi}{r}\biggl\{\int _0^r\text{d}r_2r_2\int _{r-r_2}^{r+r_2}\text{d}r_{12}r_{12}\psi(r,r_2,r_{12}) ^2\\
        &+\int _r^{r_0}\text{d}r_2r_2\int _{r_2-r}^{r_2+r}\text{d}r_{12}r_{12}\psi(r,r_2,r_{12}) ^2\biggr\}, i=1,..,4
    \end{split}
\end{equation}
where: $\psi(r,r_2,r_{12})$ is given by the equations \eqref{eqpsi1}, \eqref{eqpsi2}, \eqref{eqpsi3} or \eqref{eqpsi4} respectively.\\

The one electron probability density is normalized to unity as
\begin{equation}
    \int \rho_i(\Vec{r})d \Vec{r}=1.
\end{equation}
%==================================================================================
\subsection{Informational approach}
%==================================================================================
\subsubsection*{Shannon Entropy}
%, Aquino2013, Romera2004, Abdel2020,Lamberti2003}
The Shannon entropy \cite{Salazar2021,Shannon1948,Abdel2020,Angulo2010,Lamberti2003} is a functional of the probability density $\rho(\vec{r}) $ defined by:

\begin{equation}\label{eq:EntropyS}
    S_r=-\int \rho(\vec{r})\ln{\rho(\vec{r})}\text{d}^3\vec{r}.
\end{equation}

It quantifies the total extent of the density, it has also been used as a measure of localization-delocalization of the electron. A smaller value of $S$ corresponds to a more concentrated distribution, i. e., the particle (electron) is more localized.
%============================================================================
\subsubsection*{Kullback-Leibler entropy}
%============================================================================
The Kullback-Leibler entropy \cite{KulbackL1951,Ho1998,Majtey,Angulo2010} for a continuous probability distribution $\rho(\Vec{r})$, relative to a reference distribution $\rho_{ref}(\Vec{r})$
is defined as follows:

\begin{equation}
    KL(\rho,\rho_{ref})=\int \rho(\vec{r})\ln \frac{\rho(\vec{r})}{\rho_{ref}(\vec{r})}\text{d}\vec{r}
\end{equation}

where 

\begin{equation}
    \int \rho(\vec{r})\text{d}\vec{r}=\int \rho_{ref}(\vec{r})\text{d}\vec{r}=1,
\end{equation}

\noindent in addition $KL(\rho,\rho_{ref})\geq 0$. It can be seen that $\rho(\vec{r})=\rho_{ref}(\vec{r})\Leftrightarrow KL(\rho,\rho_{ref})=0.$
%============================================================================
\subsubsection*{Disequilibrium}
%============================================================================
Similarly the disequilibrium \cite{Estanon2020,LopezRuiz2012,LopezRuiz2015} gives us a measure between two distributions, only in this case the deviation is with respect to the equiprobability, also known as equilibrium state, it is determined as follows:

\begin{equation}
    D=\int \rho^2 (\vec{r})\text{d}\vec{r}.
\end{equation}
%=========================================================================
\subsubsection*{Tsallis entropy}
%=========================================================================

In this section the Tsallis entropy for the confined helium atom is studied using a wave function with electronic correlation in order to obtain a measure of the correlation intensity. The Tsallis entropy \cite{Nasser2021,Vershynina2023, Tsallis1998,Antolin2010,Angulo2011,Martins2008,Tsallis2009} is defined as follows:
\begin{equation}
    S_q\equiv \frac{1}{q-1}\left(1-\int  \rho^q(\vec{r})\text{d}\vec{r}\right).
\end{equation}

The Tsallis index $q$ plays a crucial role in identifying the magnitude of correlations in a system. The q value is around 1, but $q \neq 1$, in a correlated system. In the limit $q \rightarrow 1$, $ S_\textit{q} \rightarrow S_r $, i. e. the Shannon entropy is recovered. 
%=================================================
\subsubsection*{Fisher-Shannon complexity}
%=================================================
Fisher-Shannon complexity measure for a probability density $\rho$ is defined jointly by the Fisher information $F_r[\rho]$ and the Shannon entropic power. The Fisher information \cite{Fisher1925,Gonzalez,Nale,Sheila,Romera2004,Angulo2008,Adan,Abdel2020,Estanon2020,Estanon2021} is a point-to-point measure of the electron cloud distribution since it is a gradient functional of $\rho(\vec r)$ and in configuration space is tightly connected to the kinetic energy due to its dependence on the gradient of the distribution. It is interpreted as a measure of the tendency toward disorder, meaning that the larger this quantity is, the more ordered the distribution will be. It is defined by:

\begin{equation}
    F_r[\rho]=\int \frac{\lvert \vec{\nabla}\rho(\vec{r}) \rvert^2}{\rho(\vec{r})}
        \text{d}\vec{r}.
\end{equation}

The entropic Shannon power \cite{LopezRuiz2015} guarantees the positivity of this quantity and is defined as follows:

\begin{equation}
    J[\rho]=\frac{1}{2\pi e} e^{2S[\rho]/3} .
\end{equation}

It is common to define Fisher-Shannon complexity \cite{Romera2004,Angulo2008,Vignat2003} as follows:

\begin{equation}
    C_{FS}[\rho]=F_r[\rho]\times J[\rho]=\frac{1}{2\pi e}F[\rho] e^{2S[\rho]/3} .
    \label{eq:CFS}
\end{equation}

As a consequence of Stam's inequality \cite{Stam1959} this quantity satisfies the following inequality %\cite{33-34}:

\begin{equation}
\frac{1}{3}C_{FS}[\rho]\ge 1
\end{equation}

for any continuously differentiable probability density $\rho$. Moreover, this complexity measure is invariant under scaling transformations and translations, and is a monotone measure \cite{Rudnicki2016}.

%=========================================================================
\section{Results and discussion}
%=========================================================================

%=========================================================================
\subsubsection*{Shannon entropy}
%=========================================================================

\begin{figure}[h!]
    \centering
    \includegraphics[width=0.75\textwidth]{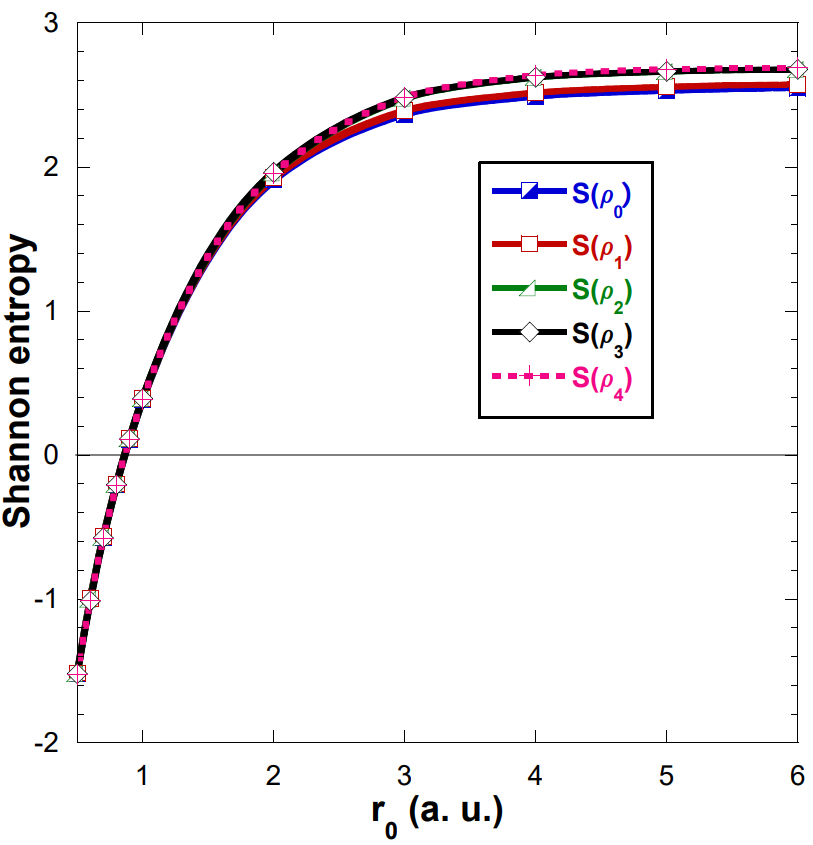}
    \caption{Shannon entropy for the helium atom confined in an impenetrable cavity with and without electronic correlation. Where $S_i=S(\rho_i)$, see Table \ref{tabla:SrCCa1}, where $i=0, 1, 2, 3, 4$. }
    \label{Fig:Sj}
  \end{figure}

From Table \ref{tabla:SrCCa1} and Figure \ref{Fig:Sj} we can see that in the confinement regime $r_0>1$  a.u. the value of the Shannon entropy for the uncorrelated density $S(\rho_0 )$, is smaller than the value of the entropies $S(\rho_i )$,\{i=1,2,3,4\}, corresponding to the functions including electronic correlation.

Gadre et. al. \cite{Gadre1985} and Hô et. al. \cite{MHo1994} used the Shannon entropy as a measure of the quality of the basis set of a free molecular system. They constructed a wave function as an expansion in a certain basis, observed that increasing the number of basis functions resulted in a better wave function, and that the Shannon entropy increased as the quality of the wave function improved. Extending Gadre's conjecture to the wave functions used in this work, we can conclude that by increasing the number of Hylleraas functions the quality of the wave function improves, i.e. it gets closer to the exact wave function.

\begin{figure}[h!]
    \centering
    \includegraphics[width=0.75\textwidth]{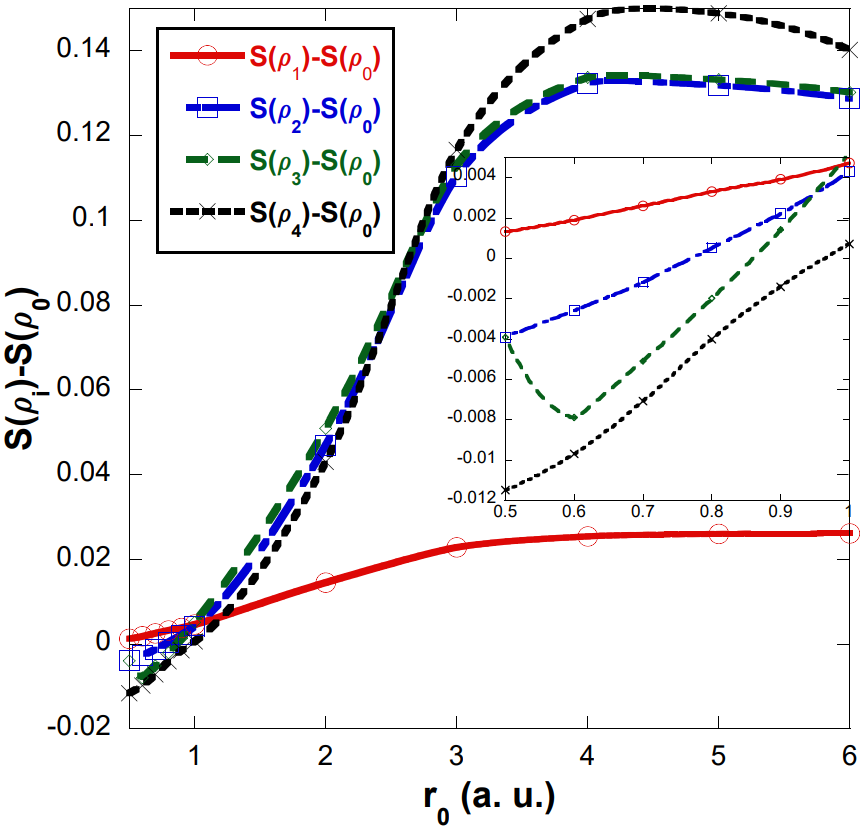}
    \caption{Shannon entropy difference $S(\rho_i)-S(\rho_0)$, where $i=1,2,3, 4$.}
    \label{Fig:Sj-s0}
  \end{figure}

Also from Figure \ref{Fig:Sj-s0} we can observe that the Shannon entropy values calculated with the electronically correlated wave functions are higher than the Shannon entropy of the uncorrelated wave function, the more correlation the wave function contains the higher the value of the Shannon entropy. This is more evident for $r_0$ greater than 2 a.u. Romera and Dehesa \cite{Romera2004} point out that this is because electronic correlation produces a dispersion of the electronic cloud, and therefore, the Shannon entropy increases.

An entirely different situation occurs in the strong confinement regime where $r_0<1$ a.u.. The value of the Shannon entropy for the uncorrelated wave function $S(\rho_0 )$,  is smaller than $S(\rho_1 )$, the Shannon entropy associated with $\psi_1$, but is larger than the entropy values for the other correlated wave functions. If Gadre's conjecture could be applied to this situation we would conclude that the best wave function, of those used in this report, would be $\psi_1$,  and we could state that electronic correlation produces a spread in the electronic probability density. We should mention that our calculation of 
 the Shannon entropy, using the $\psi_1$ function, are in complete agreement with previously published results \cite{Nascimento2020}.
The other wave functions, with higher correlation content, would make the electron density more compact, contrary to what is expected. It has been shown \cite{Antonio2010, Ed2010} that in the strong confinement regime the electron kinetic energy is so large that the problem can be reasonably well treated by perturbation theory using uncorrelated wave functions, i.e. the electron correlation is not so important in this regime. Therefore, a good description of the wave function is obtained by the $\psi_1$ function.

\begin{table}[h!]
\caption{Shannon entropy for different probability densities as a function of the confining radius $r_0$, and its comparison with those reported by Sen \cite{KDSen2005}.}
\centering
 \begin{tabular}{ c c c c c c c} 
\hline
$r_0$(a.u.) & $S(\rho_0)$ & $S(\rho_1)$ & $S(\rho_2)$  & $S(\rho_3)$ & $S(\rho_{4})$ & ref. \cite{KDSen2005}\\ [0.5ex]
 \hline\hline
0.5000  & -1.5142 & -1.5129 & -1.5181 & -1.5181 & -1.5257 & \\  
0.6000  & -0.9986 & -0.9967 & -1.0012 & -1.0065 & -1.0083 & \\  
0.7000  & -0.5696 & -0.5670 & -0.5708 & -0.5747 & -0.5767 & \\  
0.8000  & -0.2046 & -0.2013 & -0.2041 & -0.2066 & -0.2086 & \\  
0.9000  & 0.1109  &  0.1148 &  0.1131 &  0.1123 &  0.1095 & 0.1515 \\  
1.0000  & 0.3867  &  0.3914 &  0.3910 &  0.3919 &  0.3874 & \\%%%%
2.0000  & 1.9117  &  1.9263 &  1.9587 &  1.9627 &  1.9548 & 2.0097  \\
3.0000  & 2.3673  &  2.3902 &  2.4777 &  2.4803 &  2.4839 & 2.5241   \\
4.0000  & 2.4906  &  2.5161 &  2.6229 &  2.6243 &  2.6381 & 2.6197   \\  
5.0000  & 2.5310  &  2.5571 &  2.6628 &  2.6642 &  2.6798 & 2.6651   \\  
6.0000  & 2.5481  &  2.5743 &  2.6768 &  2.6783 &  2.6883 & 2.7042   \\  
10.0000 & 2.5673  &  2.5937 &  2.6900 &  2.6919 &  2.7029 & 2.7106   \\  
$\infty$ & 2.5749   & 2.60159 & 2.6945 & 2.6967 &  2.7035 & 2.7117 \\[1ex]  
 \hline
\end{tabular}
\label{tabla:SrCCa1}
\end{table}

%=========================================================================
\subsubsection*{Fisher Information}
\begin{figure}[h!]
    \centering
    \includegraphics[width=0.75\textwidth]{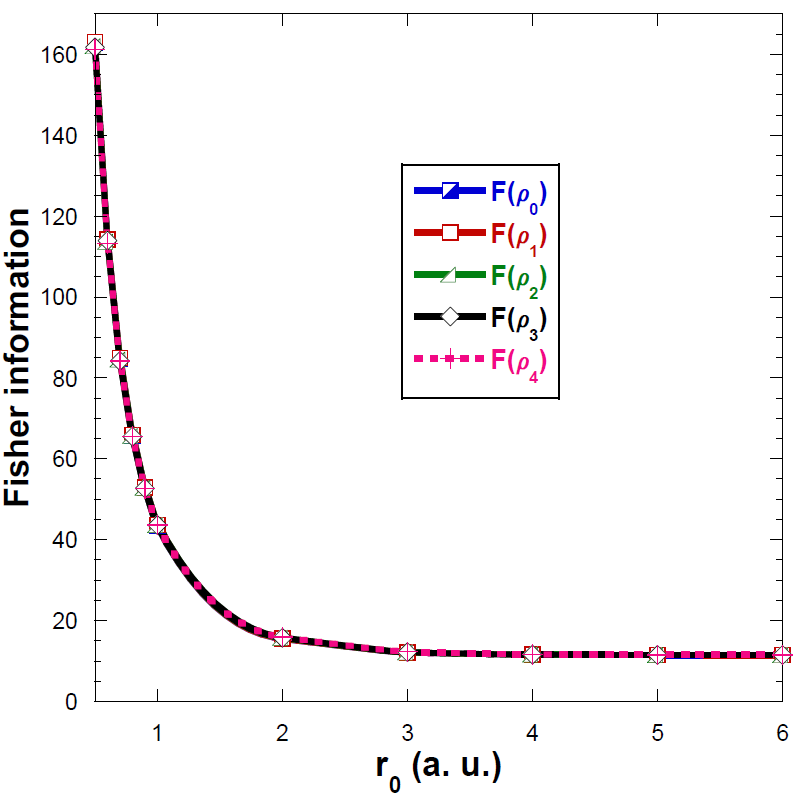}
    \caption{Fisher information for the helium atom confined in an impenetrable cavity with and without electronic correlation.}
    \label{Fig:Fj}
\end{figure}

%=========================================================================
This quantity is a measure of the concentration of the probability density. Fisher information is a local measure, which is very sensitive to variations of the probability density, even in small-sized regions. However, contrary to Shannon entropy, Fisher information decreases as $r_0$ increases, as shown in the Figure \ref{Fig:Fj}, indicating greater delocalization as $r_0$ increases. The values of Fisher information as a function of $r_0$, for the different wave functions, with and without correlation, are very similar.

\begin{figure}[h!]
    \centering
    \includegraphics[width=0.75\textwidth]{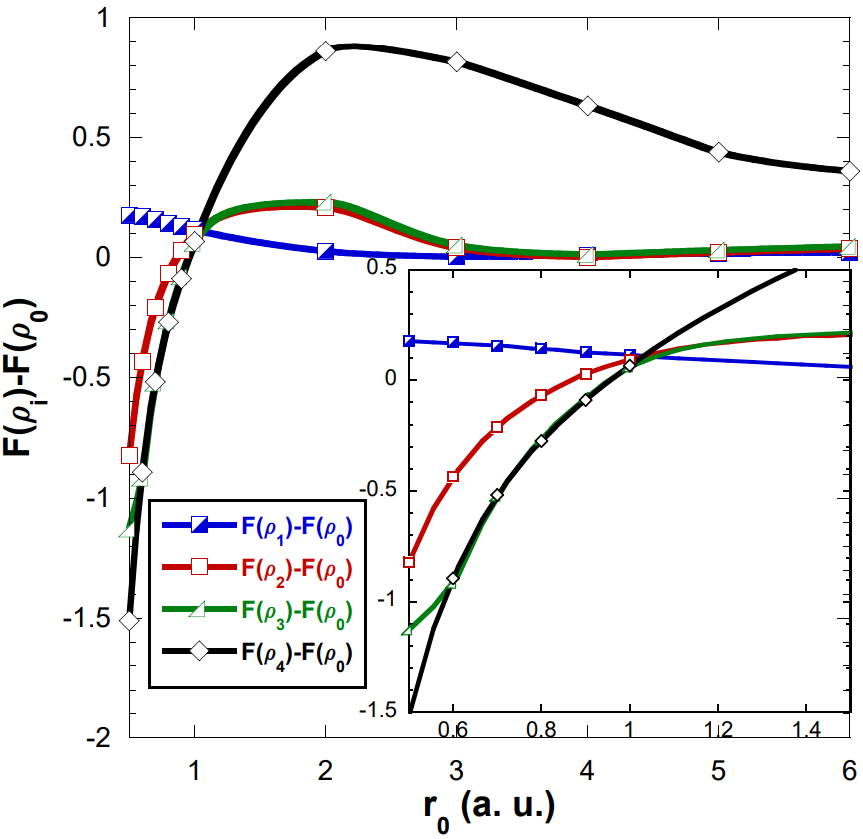}
    \caption{Fisher information for the helium atom confined in an impenetrable cavity with and without electronic correlation.}
    \label{Fig:Fj-F0}
\end{figure}

In the region $r_0>1$ a.u., the Fisher information values for the correlated wave functions are larger than the corresponding value of the Fisher information for the uncorrelated wave function.
In the strong confinement regime $r_0<1$ a.u., the Fisher information corresponding to the uncorrelated wave function $F(\rho_0 )$, is smaller than $F(\rho_1 )$. However, $F(\rho_0 )$,  is larger than the Fisher information for the $\psi_2,\psi_3$ and $\psi_4$ wave functions, which contain more electron correlation than $\psi_1$. This behavior is most evident from Figure \ref{Fig:Fj-F0} where the difference of the Fisher information for the correlated functions and the Fisher information for the uncorrelated wave function is shown. The difference between these values is entirely due to electronic correlation. It can be seen from the graph that there is a well defined maximum value around $r_0=2$ a.u., for the Fisher curves with higher correlation.

%=========================================================================
\subsubsection*{Kulback-Leibler entropy}
\begin{figure}[h!]
    \centering
    \includegraphics[width=0.8\textwidth]{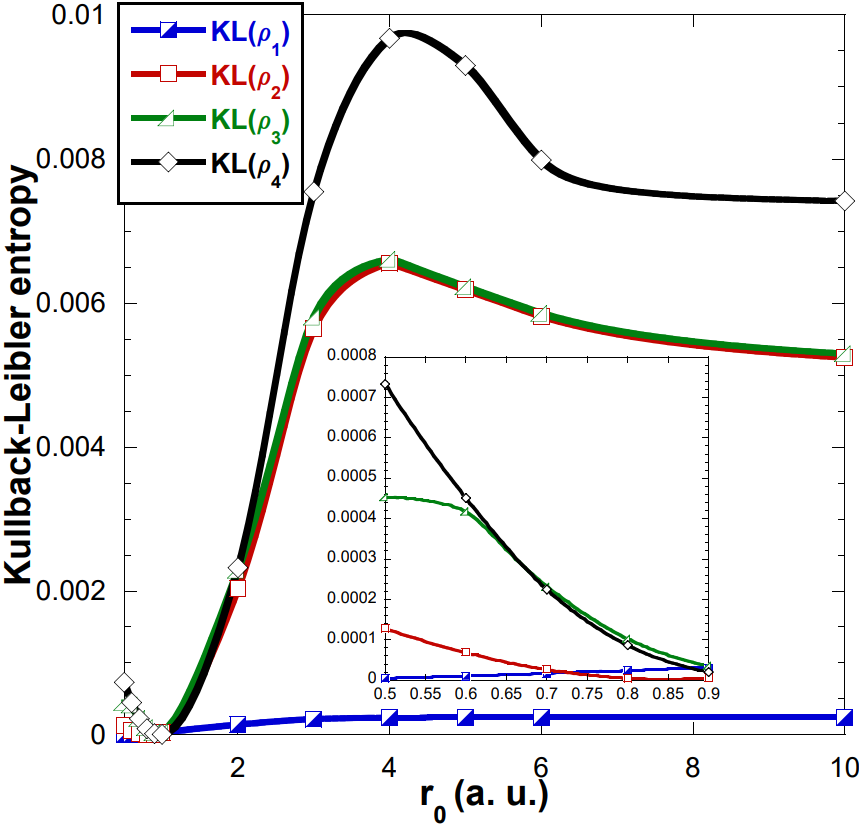}
    \caption{Kullback-Leibler entropy varying the confinement radii $r_0$ ($KL_i$ with $i=$1,2,3 and 4).}
    \label{Fig:KLD}
\end{figure}

%=========================================================================
The Kullback-Leibler (KL) entropy is a measure of the information that quantifies the amount of information with by which the probability density $\rho(\vec{r})$  differs from the reference density $\rho_0 (\vec{r})$. This measure is zero when the probability density $\rho(\vec{r})$  is identical to the reference probability density $\rho_0 (\vec{r})$. In other words, this measure quantifies the similarity between the two probability densities. When the KL entropy is small the probability densities $\rho(\vec{r})$ and $\rho_0 (\vec{r})$ are similar, and when the KL entropy is large, the two probability densities are remarkably different.
Figure \ref{Fig:KLD} shows the KL entropy values for the electronically correlated $\rho_i (\vec{r})$ densities with respect to the uncorrelated $\rho_0 (\vec{r})$ reference density. Those values are entirely due to the electronic correlation.

For values of $r_0>1$ a.u., the KL entropy values increase with $r_0$, but even so the densities $\rho_1 (\vec{r})$  and $\rho_0 (\vec{r})$ remain very similar. The KL entropies for the densities with greater correlation increase with $r_0$, and have a maximum value near $r_0=4$ a.u., and then decrease and tend asymptotically to the values of the free case. In other words the Kullback-Liebler entropy varies with $r_0$ and its highest value is found around $r_0=4$ a.u..

For $r_0<1$ a.u., the KL entropy for $\rho_1 (\vec{r})$ diminishes as $r_0$ decreases. The KL entropies for $\rho_i (\vec{r})$,  $i=2-4$, decrease as $r_0$  decreases, and reach a minimum value and increase again, indicating that the correlation decreases with $r_0$, reaches a minimum value at $\thicksim r_0=1$ a.u. and increases again.

%=========================================================================
\subsubsection*{Tsallis entropy}
\begin{figure}[h!]
    \centering
    \includegraphics[width=0.75\textwidth]{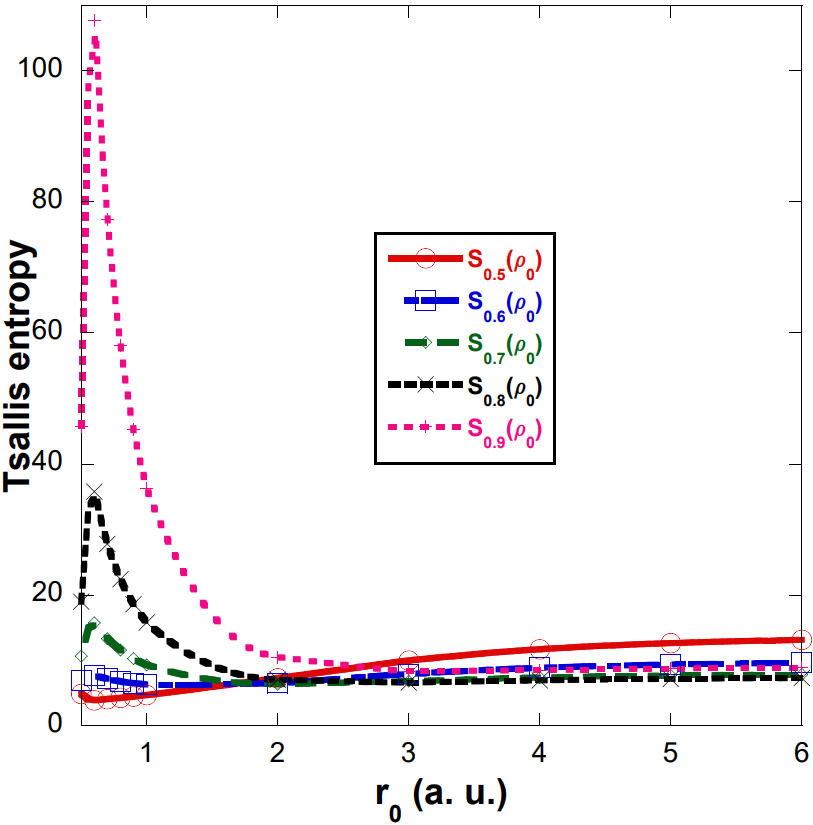}
    \caption{Tsallis entropy for the helium atom confined in an impenetrable cavity without electronic correlation for $\rho_3$.}
    \label{Fig:Tsrho0}
\end{figure}

%=========================================================================
The analysis shown below is for $\rho_3 (\vec{r})$, however, the behavior of Tsallis entropy for $\rho_i (\vec{r})$,   ($i=1, 2$ and $4$) is very similar.
The Tsallis entropy with a \textit{q} value different from 1 has been used as a measure of the electronic correlation \cite{Nasser2021}. Figure \ref{Fig:Tsrho0} shows the Tsallis entropy using a wave function without electronic correlation, where values of \textit{q} = 0.5,...,0.9 were used. It should be noted that this plot does not provide correlation information. The Tsallis entropy curves have a maximum around 0.6 a.u. and this becomes more pronounced as q approaches 1. We must remember that in the limiting case $\textit{q} \rightarrow 1$, Tsallis entropy becomes the Shannon entropy, shown in Figures 2 and 3.

\begin{figure}[h!]
    \centering
    \includegraphics[width=0.75\textwidth]{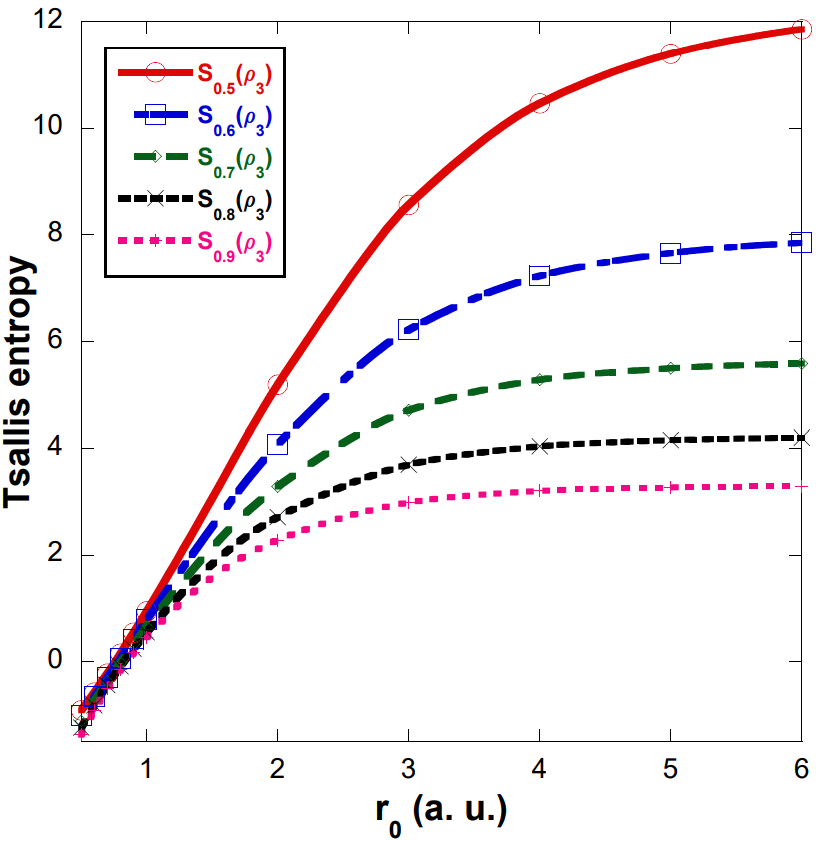}
    \caption{Tsallis entropy for the helium atom confined in an impenetrable cavity with electronic correlation for $\rho_3$.}
    \label{Fig:Tsrho3}
\end{figure}

\begin{figure}[h!]
    \centering
    \includegraphics[width=0.75\textwidth]{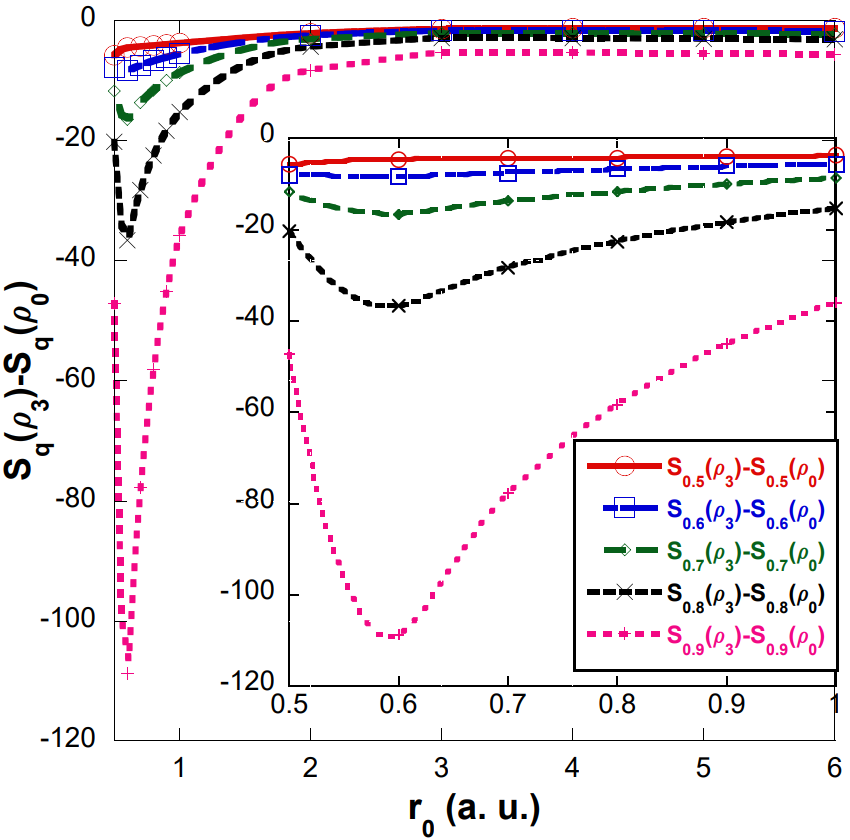}
    \caption{Tsallis entropy for the helium atom confined in an impenetrable cavity with $S_q(\rho_3)-S_q(\rho_0)$}
    \label{Fig:Ts3-Ts0}
\end{figure}

On the other hand, Figure \ref{Fig:Tsrho3} shows the Tsallis entropy with electronic correlation for the same values of \textit{q}, here we can notice that as the confinement radius becomes smaller the correlation decreases, as expected, since the kinetic energy is greater than the potential energy in the region of strong confinement. In addition, we notice that the correlation is greater for \textit{q} = 0.5 and as we increase the value of q the correlation decreases.  Finally, in Figure \ref{Fig:Ts3-Ts0} we plot the difference: Tsallis entropy with correlation - Tsallis entropy without correlation. On the other hand, the difference $S_q (\rho_3 )-S_q (\rho_0 )$ is due completely to the correlation. 

%Those quantities shows a similar behavior to correlation energy \cite{Wilson2010}. In this way it would indicate that this entropy is giving us information not only about the correlation but also about the correlation energy. \textcolor{red}{Montgomery indica que no entiende la afirmación sobre la energía de correlación, se podría indicar la comparación como se hizo en la complejidad de Fisher-Shannon}.

%=========================================================================
\subsubsection*{Fisher-Shannon complexity}
\begin{figure}[h!]
    \centering
    \includegraphics[width=0.75\textwidth]{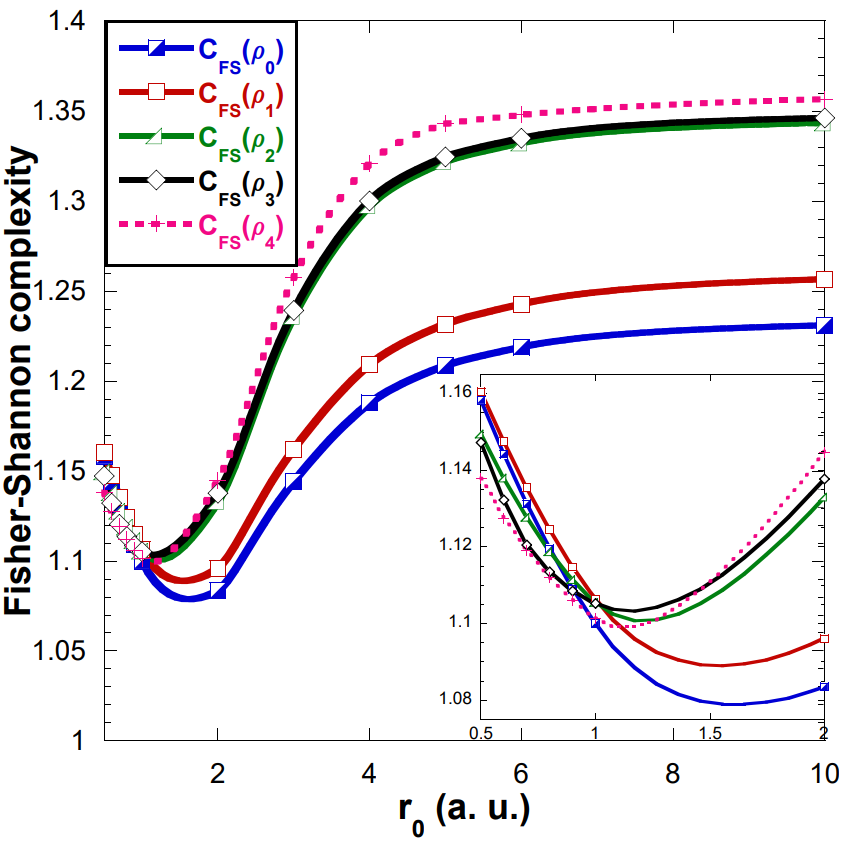}
    \caption{Fisher-Shannon complexity for the helium atom confined in an impenetrable cavity with and without electronic correlation.}
    \label{Fig:CFS}
\end{figure}

%=========================================================================
The Fisher-Shannon complexity is a measure of the probability density distribution in a global-local form that has been used as a measure of the correlation energy by Dehesa et. al. \cite{Romera2004}. This interpretation makes sense if we look at Figure \ref{Fig:CFS}, in which we notice that around $r_0=1.5$ a.u. there is a minimum, which is close to the value of $r_0=2$ a.u., at which the maximum of the correlation energy was found by Wilson et. al. \cite{Wilson2010}.
For values of $r_0<1$ a.u. the values of all curves are very similar. For $r_0>1$ a.u. the Fisher-Shannon complexities for the correlated wave functions are higher than for the uncorrelated wave function.

%=========================================================================
\section{Conclusions}
%=========================================================================

In this work we obtained the energies and wave functions of the helium atom confined in a spherical box with impenetrable walls. We used the variational method and as trial wave functions we employed one function without electronic correlation ($\psi_0$, Eq.(4)) and four functions with different degree of correlation (Eqs. (5)-(8)). We constructed the probability densities of the five test functions and calculated the Shannon entropy, Fisher information, Kullback-Leibler entropy, Tsallis entropy and Fisher-Shannon complexity as a function of the radius of the spherical box $r_0$.\newline
We find that the behavior of all information measures, used in this work, is different in the regions $r_0<1$ a.u. (strong confinement region) and $r_0>1$ a.u.. For each of the information measures, the difference between the values obtained with a wave function with correlation $\psi_i$, (i=2-4) and with the uncorrelated function $\psi_0$, is entirely due to electronic correlation. All information measures show evidence of electronic correlation. The electronic correlation is not constant but varies with $r_0$. The maximum value of the correlation measure varies with the type of information measure used. For example, for Shannon entropy this maximum value is reached around $r_0=4$ a.u., while for Fisher information at $r_0=2$ a.u., for Kullback-Leibler entropy at $r_0=4$ a.u., while for the Tsallis entropy a minimum is observed around $r_0=0.6$ a.u., and for the Fisher-Shannon complexity the minimum is located around $r_0=1.6$ a.u., which is close to the point $r_0=2$ a.u., at which Wilson et al. \cite{Wilson2010} find a higher correlation energy.\newline

\section*{Acknowledgements}

N.A. and C. E. would like to thank R. P. Sagar for his comment on an early version of this work. J.C.A. belongs to the research group FQM-207, and gratefully acknowledges financial support by the Spanish projects PID2020-113390GB-I00 (MICIN), PY20 00082 (ERDF-Junta de Andalucía), and AFQM-52-UGR20 (ERDF-University of Granada).
\clearpage

\bibliography{Referencias.bib}

\begin{thebibliography}{10}

\bibitem{Froman87}
P.~O. Fr{\"o}man, S.~Yngve, and N.~Fr{\"o}man.
\newblock The energy levels and the corresponding normalized wave functions for
  a model of a compressed atom.
\newblock {\em J. Math. Phys.}, 28(8):1813--1826, 1987.

\bibitem{Jasko96}
W.~Jask{\'o}lski.
\newblock Confined many-electron systems.
\newblock {\em Phys. Rep.}, 271(1):1--66, 1996.

\bibitem{Buch}
A.~L. Buchachenko.
\newblock Compressed atoms.
\newblock {\em J. Phys. Chem. B}, 105(25):5839--5846, 2001.

\bibitem{Conne}
J.~P. Connerade, V.~K. Dolmatov, and P.~A. Lakshmi.
\newblock The filling of shells in compressed atoms.
\newblock {\em J. Phys. B: At. Mol. Opt. Phys.}, 33(2):251, 2000.

\bibitem{Sabin2009}
J.~R. Sabin, E.~Br{\"a}ndas, and S.~A. Cruz, editors.
\newblock {\em Theory of Confined Quantum Systems}.
\newblock Academic Press, Parts I and II, 2009.

\bibitem{Sen2009}
K.~D. Sen, V.~I. Pupyshev, and Montgomery~Jr. HE.
\newblock Exact relations for confined one-electron systems.
\newblock {\em Adv. Quantum Chem.}, 57:25--77, 2009.

\bibitem{Sen2014}
K.~D. Sen, editor.
\newblock {\em Electronic structure of quantum confined atoms and molecules}.
\newblock Springer, Switzerland, 2014.

\bibitem{LeyKoo2018}
E.~Ley-Koo.
\newblock Recent progress in confined atoms and molecules: Superintegrability
  and symmetry breakings.
\newblock {\em Rev. M{\'e}x. Fis.}, 64(4):326--363, 2018.

\bibitem{Banyai}
S.~W. Koch.
\newblock {\em Semiconductor Quantum Dots}, volume~2.
\newblock World Scientific, 1993.

\bibitem{Harrison}
P.~Harrison and A.~Valavanis.
\newblock {\em Quantum Wells, Wires and Dots: Theoretical and Computational
  Physics of Semiconductor Nanostructures}.
\newblock John Wiley \& Sons, 2016.

\bibitem{Michels}
A.~Michels, J.~De~Boer, and A.~Bijl.
\newblock Remarks concerning molecular interaction and their influence on the
  polarisability.
\newblock {\em Physica}, 4(10):981--994, 1937.

\bibitem{Ley1}
E.~Ley-Koo and S.~Rubinstein.
\newblock The hydrogen atom within spherical boxes with penetrable walls.
\newblock {\em J. Chem. Phys.}, 71(1):351--357, 1979.

\bibitem{Marin}
J.~L. Marin and S.~A. Cruz.
\newblock Use of the direct variational method for the study of one-and
  two-electron atomic systems confined by spherical penetrable boxes.
\newblock {\em J. Phys. B: At. Mol. Opt. Phys.}, 25(21):4365, 1992.

\bibitem{Aquino2009}
N.~Aquino.
\newblock The hydrogen and helium atoms confined in spherical boxes.
\newblock {\em Adv. Quantum Chem.}, 57:123--171, 2009.

\bibitem{Sen2}
KD~Sen, VI~Pupyshev, and HE~Montgomery~Jr.
\newblock Exact relations for confined one-electron systems.
\newblock {\em Adv. Quantum Chem.}, 57:25--77, 2009.

\bibitem{Ed1}
Montgomery~Jr. HE and K.~D. Sen.
\newblock Dipole polarizabilities for a hydrogen atom confined in a penetrable
  sphere.
\newblock {\em Phys. Lett. A}, 376(26-27):1992--1996, 2012.

\bibitem{tenseldam}
C.~A. Ten~Seldam and S.~R. De~Groot.
\newblock On the ground state of a model for compressed helium.
\newblock {\em Physica}, 18(11):891--904, 1952.

\bibitem{Gimarc}
B.~M. Gimarc.
\newblock Correlation energy of the two-electron atom in a spherical potential
  box.
\newblock {\em J. Chem. Phys.}, 47(12):5110--5115, 1967.

\bibitem{Lude1}
E.~V. Ludeña.
\newblock Scf hartree–fock calculations of ground state wavefunctions of
  compressed atoms.
\newblock {\em J. Chem. Phys.}, 69(4):1770--1775, 1978.

\bibitem{Luden2}
E.~V. Ludeña and M.~Gregori.
\newblock Configuration interaction calculations for two-electron atoms in a
  spherical box.
\newblock {\em J. Chem. Phys.}, 71(5):2235--2240, 1979.

\bibitem{Joslin}
C.~Joslin and S.~Goldman.
\newblock Quantum montecarlo studies of two-electron atoms constrained in
  spherical boxes.
\newblock {\em J. Phys. B: At. Mol. Opt. Phys.}, 25(9):1965, 1992.

\bibitem{Aquino2003}
N.~Aquino, A.~Flores-Riveros, and J.~F. Rivas-Silva.
\newblock The compressed helium atom variationally treated via a correlated
  \uppercase{H}ylleraas wave function.
\newblock {\em Phys. Lett. A}, 307(5-6):326--336, 2003.

\bibitem{Aquino2006}
N.~Aquino, J.~Garza, A.~Flores-Riveros, J.~F. Rivas-Silva, and K.~D. Sen.
\newblock Confined helium atom low-lying s states analyzed through correlated
  \uppercase{H}ylleraas wave functions and the
  \uppercase{K}kohn-\uppercase{S}ham model.
\newblock {\em J. Chem. Phys.}, 124(5):054311, 2006.

\bibitem{Antonio2008}
A.~Flores-Riveros and A.~Rodriguez-Contreras.
\newblock Compression effects in helium-like atoms (\uppercase{Z}= 1,..., 5)
  constrained by hard spherical walls.
\newblock {\em Phys. Lett. A}, 372(40):6175--6182, 2008.

\bibitem{Antonio2010}
A.~Flores-Riveros, N.~Aquino, and Montgomery~Jr. HE.
\newblock Spherically compressed helium atom described by perturbative and
  variational methods.
\newblock {\em Phys. Lett. A}, 374(10):1246--1252, 2010.

\bibitem{Laughlin}
C.~Laughlin and S.~I. Chu.
\newblock A highly accurate study of a helium atom under pressure.
\newblock {\em J. Phys. A: Theor. Math. Phys.}, 42(26):265004, 2009.

\bibitem{Ed2010}
Montgomery~Jr. HE, N.~Aquino, and A.~Flores-Riveros.
\newblock The ground state energy of a helium atom under strong confinement.
\newblock {\em Phys. Lett. A}, 374(19-20):2044--2047, 2010.

\bibitem{Wilson2010}
C.~L. Wilson, Montgomery~Jr. HE, K.~D. Sen, and D.~C. Thompson.
\newblock Electron correlation energy in confined two-electron systems.
\newblock {\em Phys. Lett. A}, 374(43):4415--4419, 2010.

\bibitem{Aquino2014}
N.~Aquino.
\newblock The role of correlation in the ground state energy of confined helium
  atom.
\newblock In {\em AIP Conference Proceedings}, volume 1579, pages 136--140,
  2014.

\bibitem{Lesech}
C.~Le~Sech and A.~Banerjee.
\newblock A variational approach to the \uppercase{D}irichlet boundary
  condition: application to confined h-, he and li.
\newblock {\em J. Phys. B: At. Mol. Opt. Phys.}, 44(10):105003, 2011.

\bibitem{Bhatta}
S.~Bhattacharyya, J.~K. Saha, P.~K. Mukherjee, and T.~K. Mukherjee.
\newblock Precise estimation of the energy levels of two-electron atoms under
  spherical confinement.
\newblock {\em Physica Scripta}, 87(6):065305, 2013.

\bibitem{Jgo2016}
T.~D. Young, R.~Vargas, and J.~Garza.
\newblock A \uppercase{H}artree-\uppercase{F}ock study of the confined helium
  atom: Local and global basis set approaches.
\newblock {\em Phys. Lett. A}, 380(5-6):712--717, 2016.

\bibitem{Doma}
S.~B. Doma and F.~N. El-Gammal.
\newblock Application of variational montecarlo method to the confined helium
  atom.
\newblock {\em J. Theor. Appl. Phys.}, 6(1):1--7, 2012.

\bibitem{Ben2006}
A.~Ben~Hamza.
\newblock Nonextensive information--theoretic measure for image edge detection.
\newblock {\em Journal of Electronic Imaging}, 15(1):013011--013011, 2006.

\bibitem{Martins2008}
A.~F. Martins, P.~M. Aguiar, and M.~A. Figueiredo.
\newblock \uppercase{T}sallis kernels on measures.
\newblock In {\em 2008 IEEE Information Theory Workshop}, pages 298--302. IEEE,
  2008.

\bibitem{Martins2009}
A.~F. Martins, N.~A. Smith, E.~P. Xing, P.~M. Aguiar, and M.~A. Figueiredo.
\newblock Nonextensive information theoretic kernels on measures.
\newblock {\em Journal of Machine Learning Research}, 10(4), 2009.

\bibitem{Salazar2021}
S.~J.~C. Salazar, H.~G. Laguna, B.~Dahiya, V.~Prasad, and R.~P. Sagar.
\newblock \uppercase{S}hannon information entropy sum of the confined
  hydrogenic atom under the influence of an electric field.
\newblock {\em The European Physical Journal D}, 75(4):127, 2021.

\bibitem{Laguna2022}
H.~G. Laguna, S.~J. Salazar, and R.~P. Sagar.
\newblock Information theoretical statistical discrimination measures for
  electronic densities.
\newblock {\em Journal of Mathematical Chemistry}, 60(7):1422--1444, 2022.

\bibitem{Salazar2022}
S.~J. Salazar, H.~G. Laguna, and R.~P. Sagar.
\newblock Pairwise and higher-order statistical correlations in excited states
  of quantum oscillator systems.
\newblock {\em The European Physical Journal Plus}, 137(1):1--26, 2022.

\bibitem{Salazar2023}
S.~J. Salazar, H.~G. Laguna, and R.~P. Sagar.
\newblock Phase-space quantum distributions and information theory.
\newblock {\em Physical Review A}, 107(4):042417, 2023.

\bibitem{Cesar}
C.~Martínez-Flores, M.~A. Martínez-Sánchez, R.~Vargas, and J.~Garza.
\newblock Free-basis-set method to describe the helium atom confined by a
  spherical box with finite and infinite potentials.
\newblock {\em Eur. Phys. J. D}, 75(3):1--9, 2021.

\bibitem{Yanes1994}
R.~J. Y\'añez, W.~V. Assche, and J.~S. Dehesa.
\newblock Position and momentum information entropies of the d-dimensional
  harmonic oscillator and hydrogen atom.
\newblock {\em Phys. Rev. A}, 50:3065--3079, 1994.

\bibitem{Sen2007}
K.~D. Sen, C.~P. Panos, K.~Ch. Chatzisavvas, and Ch.~C. Moustakidis.
\newblock Net \uppercase{F}isher information measure versus ionization
  potential and dipole polarizability in atoms.
\newblock {\em Physics Letters A}, 364(4-5):286--290, 2007.

\bibitem{Aquino2013}
N.~Aquino, A.~Flores-Riveros, and J.~F. Rivas-Silva.
\newblock \uppercase{S}hannon and \uppercase{F}isher entropies for a hydrogen
  atom under soft spherical confinement.
\newblock {\em Physics Letters A}, 377(32-33):2062--2068, 2013.

\bibitem{KDSen2011}
K.~D. Sen, editor.
\newblock {\em Statistical complexity: applications in electronic structure}.
\newblock Springer Science \& Business Media, 2011.

\bibitem{HoSmith}
M.~Ho, V.~H. Smith~Jr., D.~F. Weaver, C.~Gatti, R.~P. Sagar, and R.~O.
  Esquivel.
\newblock Molecular similarity based on information entropies and distances.
\newblock {\em Journal of Chemical Information and Modeling}, 38(5):546--558,
  1998.

\bibitem{Ho1998}
M.~Ho, D.~F. Weaver, V.~H. Smith~Jr., R.~P. Sagar, and R.~O. Esquivel.
\newblock Calculating the logarithmic mean excitacion energy from the
  \uppercase{S}hannon information entropy of the electronic charge density.
\newblock {\em Phys. Rev. A}, 57:4512--4517, 1998.

\bibitem{Gadre1984}
S.~R. Gadre.
\newblock Information entropy and \uppercase{T}homas-\uppercase{F}ermi theory.
\newblock {\em Phys. Rev. A}, 30:620--621, 1984.

\bibitem{Gadre1985}
S.~R. Gadre, S.~B. Sears, S.~J. Chakravorty, and R.~D. Bendale.
\newblock Some novel characteristics of atomic information entropies.
\newblock {\em Phys. Rev. A}, 32:2602, 1985.

\bibitem{Gadre1985b}
S.~R. Gadre, R.~J. Bendale, and A.~P. Gejji.
\newblock Analysis of atomic electron momentum densities: use of information
  entropies in coordinate and momentum space.
\newblock {\em J. Phys. B: At. Mol. Phys.}, 18:138--142, 1985.

\bibitem{Pan2003}
X.~Y. Pan, V.~Sahni, L.~Massa, and K.~D. Sen.
\newblock New expression for the expectation value integral for a confined
  helium atom.
\newblock {\em J. Theor. Comput. Chem.}, 965(1):202--205, 2011.

\bibitem{Nascimento2020}
W.~S. Nascimento, M.~M. de~Almeida, and F.~V. Prudente.
\newblock Coulomb correlation and information entropies in confined helium-like
  atoms.
\newblock {\em Eur. Phys. J. D}, 2021.

\bibitem{Shannon1948}
C.~E. Shannon.
\newblock A mathematical theory of communication.
\newblock {\em The Bell System Technical Journal}, 27(3):379--423, 1948.

\bibitem{Abdel2020}
I.~Nasser and A.~Abdel-Hady.
\newblock \uppercase{F}isher information and \uppercase{S}hannon entropy
  calculations for two-electron systems.
\newblock {\em Canadian Journal of Physics}, 98(8):784--789, 2020.

\bibitem{Angulo2010}
J.~C. Angulo, J.~Antol\'in, S.~L\'opez-Rosa, and R.~O. Esquivel.
\newblock Jensen–\uppercase{S}hannon divergence in conjugate spaces: The
  entropy excess of atomic systems and sets with respect to their constituents.
\newblock {\em Physica A: Statistical Mechanics and its Applications},
  389(4):899--907, 2010.

\bibitem{Lamberti2003}
P.~W. Lamberti and A.~P. Majtey.
\newblock Non-logarithmic jensen--\uppercase{S}hannon divergence.
\newblock {\em Physica A: Statistical Mechanics and its Applications},
  329(1--2):81--90, 2003.

\bibitem{KulbackL1951}
S.~Kullback and R.~A. Leibler.
\newblock On information and sufficiency.
\newblock {\em The Annals of Mathematical Statistics}, 22(1):79--86, 1951.

\bibitem{Majtey}
A~Majtey, Pedro~W Lamberti, Mar{\'\i}a~Teresa Martin, and A~Plastino.
\newblock Wootters’ distance revisited: a new distinguishability criterium.
\newblock {\em The European Physical Journal D-Atomic, Molecular, Optical and
  Plasma Physics}, 32:413--419, 2005.

\bibitem{Estanon2020}
C.~R. Estañón, N.~Aquino, D.~Puertas-Centeno, and J.~S. Dehesa.
\newblock Two-dimensional confined hydrogen: An entropy and complexity
  approach.
\newblock {\em Int. J. Quantum Chem.}, 120(11):e26192, 2020.

\bibitem{LopezRuiz2012}
R.~López-Ruiz, J.~Sañudo, E.~Romera, and X.~Calbet.
\newblock Statistical complexity and \uppercase{F}isher-\uppercase{S}hannon
  information: Applications.
\newblock {\em Statistical Complexity: Applications in Electronic Structure},
  pages 65--127, 2011.

\bibitem{LopezRuiz2015}
R.~López~Ruiz and J.~Sañudo.
\newblock Statistical complexity. applications in electronic systems.
\newblock {\em J. Comput. Sci.}, (ART-2015-95144), 2015.

\bibitem{Nasser2021}
I.~Nasser, C.~Martinez-Flores, M.~Zeama, R.~Vargas, and J.~Garza.
\newblock Tsallis entropy: A comparative study for the 1s2-state of helium
  atom.
\newblock {\em Phys. Lett. A}, 392:127136, 2021.

\bibitem{Vershynina2023}
A.~Vershynina.
\newblock Coherence as entropy increment for \uppercase{T}sallis and
  \uppercase{R}ényi entropies.
\newblock {\em Quantum Inf. Process.}, 22(2):127, 2023.

\bibitem{Tsallis1998}
C.~Tsallis.
\newblock Possible generalization of \uppercase{B}oltzmann-\uppercase{G}ibbs
  statistics.
\newblock {\em Journal of Statistical Physics}, 52:479--487, 1988.

\bibitem{Antolin2010}
J.~Antol\'in, S.~L\'opez-Rosa, J.~C. Angulo, and R.~O. Esquivel.
\newblock Jensen--\uppercase{T}sallis divergence and atomic dissimilarity for
  position and momentum space electron densities.
\newblock {\em The Journal of Chemical Physics}, 132(4):044105, 2010.

\bibitem{Angulo2011}
J.~C. Angulo, J.~Antol{\'\i}n, S.~L{\'o}pez-Rosa, and R.~O. Esquivel.
\newblock Jensen--\uppercase{T}sallis divergence and atomic dissimilarity for
  ionized systems in conjugated spaces.
\newblock {\em Physica A: Statistical Mechanics and its Applications},
  390(4):769--780, 2011.

\bibitem{Tsallis2009}
C.~Tsallis.
\newblock Computational applications of nonextensive statistical mechanics.
\newblock {\em Journal of Computational and Applied Mathematics},
  227(1):51--58, 2009.

\bibitem{Fisher1925}
R.~A. Fisher.
\newblock Theory of statistical estimation.
\newblock {\em Proc. Camb. Phil. Soc.}, 22:700--725, 1925.

\bibitem{Gonzalez}
R~Gonz{\'a}lez-F{\'e}rez and JS~Dehesa.
\newblock Characterization of atomic avoided crossings by means of
  \uppercase{F}isher’s information.
\newblock {\em The European Physical Journal D-Atomic, Molecular, Optical and
  Plasma Physics}, 32:39--43, 2005.

\bibitem{Nale}
R.~F. Nalewajski.
\newblock {\em Information Theory of Molecular Systems}.
\newblock Elsevier Science, New York, 2006.

\bibitem{Sheila}
Sheila L{\'o}pez-Rosa, Rodolfo~O Esquivel, Juan~Carlos Angulo, Juan
  Antol{\'\i}n, Jes{\'u}s~S Dehesa, and Nelson Flores-Gallegos.
\newblock \uppercase{F}isher information study in position and momentum spaces
  for elementary chemical reactions.
\newblock {\em Journal of Chemical Theory and Computation}, 6(1):145--154,
  2010.

\bibitem{Romera2004}
E.~Romera and J.~S. Dehesa.
\newblock The \uppercase{F}isher-\uppercase{S}hannon information plane, an
  electron correlation tool.
\newblock {\em The Journal of Chemical Physics}, 120(19):8906--8912, 2004.

\bibitem{Angulo2008}
J.~C. Angulo, J.~Antolín, and K.~D. Sen.
\newblock \uppercase{F}isher-\uppercase{S}hannon plane and statistical
  complexity of atoms.
\newblock {\em Physics Letters A}, 372(5):670--674, 2008.

\bibitem{Adan}
M.~A. Martínez-Sánchez, C.~Martínez-Flores, R.~Vargas, J.~Garza,
  R.~Cabrera-Trujillo, and K.~D. Sen.
\newblock Ionization of many-electron atoms by the action of two plasma models.
\newblock {\em Physical Review E}, 103(4):043202, 2021.

\bibitem{Estanon2021}
C.~R. Estañón, N.~Aquino, D.~Puertas-Centeno, and J.~S. Dehesa.
\newblock Cramér-\uppercase{R}ao complexity of the confined two-dimensional
  hydrogen.
\newblock {\em Int. J. Quantum Chem.}, 121(2):e26424, 2021.

\bibitem{Vignat2003}
C.~Vignat and J.~F. Bercher.
\newblock Analysis of signals in the \uppercase{F}isher-\uppercase{S}hannon
  information plane.
\newblock {\em Physics Letters A}, 312(1-2):27--33, 2003.

\bibitem{Stam1959}
A.~J. Stam.
\newblock Some inequalities satisfied by the quantities of information of
  \uppercase{F}isher and \uppercase{S}hannon.
\newblock {\em Information and Control}, 2(2):101--112, 1959.

\bibitem{Rudnicki2016}
{\L}.~Rudnicki, I.~V. Toranzo, P.~Sánchez-Moreno, and J.~S. Dehesa.
\newblock Monotone measures of statistical complexity.
\newblock {\em Phys. Lett. A}, 380(3):377--380, 2016.

\bibitem{MHo1994}
Minhhuy H{\^o}, Robin~P Sagar, Vedene~H Smith~Jr., and Rodolfo~O Esquivel.
\newblock Atomic information entropies beyond the
  \uppercase{H}artree-\uppercase{F}ock limit.
\newblock {\em Journal of Physics B: Atomic, Molecular and Optical Physics},
  27(21):5149, 1994.

\bibitem{KDSen2005}
K.~D. Sen.
\newblock Characteristic features of \uppercase{S}hannon information entropy of
  confined atoms.
\newblock {\em J. Chem. Phys.}, 123:074110, 2005.

\end{thebibliography}
\bibliographystyle{unsrt}
\nocite{*}  

\end{document}